# Title: Imaging One-dimensional Quantum Melting of Anisotropic Wigner Crystals

**Authors:** Ziyu Xiang[1, 2, 3, 6]†, Jianghan Xiao[1, 2, 3]†, Hongyuan Li[1, 2, 6], Woochang Kim [1, 3], Tianle Wang[1], Zhihuan Dong[1], Takashi Taniguchi[4], Kenji Watanabe[5], Michael P. Zaletel[1], Steven G. Louie[1, 3], Michael F. Crommie[1, 3, 6]* and Feng Wang[1, 3, 6]*

**Affiliations:**

[1]Department of Physics, University of California at Berkeley; Berkeley, CA, USA.

[2]Graduate Group in Applied Science and Technology, University of California at Berkeley; Berkeley, CA, USA.

[3]Materials Sciences Division, Lawrence Berkeley National Laboratory; Berkeley, CA, USA.

[4]Research Center for Materials Nanoarchitectonics, National Institute for Materials Science; 1-1 Namiki, Tsukuba 305-0044, Japan

[5]Research Center for Electronic and Optical Materials, National Institute for Materials Science; 1-1 Namiki, Tsukuba 305-0044, Japan

[6]Kavli Energy Nano Sciences Institute at the University of California Berkeley and the Lawrence Berkeley National Laboratory; Berkeley, CA, USA.

†These authors contributed equally: Ziyu Xiang, and Jianghan Xiao

*Corresponding author: crommie@berkeley.edu and fengwang76@berkeley.edu.

**Abstract:** We directly visualize a two-dimensional anisotropic Wigner crystal and its quantum melting in monolayer 1T'-$ReSe_2$ using non-invasive scanning tunnelling microscopy. Solids composed of anisotropic molecules can form ordered structures having reduced symmetry. Upon heating, anisotropic molecular crystals can melt into a rich variety of liquid crystal phases at intermediate temperatures before transitioning to liquid at high temperature(*1*). Electrons can also assume an "anisotropic" shape in its quantum wavefunction when the electron effective mass is anisotropic in the crystal structure: the electron wavefunction will be more spread out along the light mass direction to lower the quantum kinetic energy(*2–4*). Such anisotropic electrons are predicted to form oblique quantum crystals, also known as anisotropic Wigner crystals(*2*, *3*), at low electron densities. With increased electron density anisotropic Wigner crystals could melt into exotic electron liquid crystal phases due to quantum fluctuations(*5*). Anisotropic Wigner crystal physics, however, has been little explored experimentally. Here, we directly visualize a two-dimensional anisotropic Wigner crystal and its quantum melting in monolayer 1T'-$ReSe_2$ using non-invasive scanning tunnelling microscopy. We first demonstrated "shape anisotropy" of individual anisotropic electrons in gated monolayer $ReSe_2$, where the quantum wavefunctions of single electrons are strongly elongated shape along the light-mass direction. These elongated electrons crystallize into an oblique Wigner lattice at low electron density, which is quite different from the well-known triangular electron lattice seen for isotropic crystals. At increased electron density, quantum fluctuations in the light-mass direction increase faster than in the heavy-mass direction, resulting in one-dimensional (1D) melting of the Wigner crystal. The 1D-melted Wigner crystal can be viewed as a smectic liquid crystal state(*1*, *6*) that lies between the electron solid and Fermi liquid phases. This provides a new platform to explore coupled 1D electron chains, which might give rise to sliding Luttinger liquids(*7*), non-abelian quantum Hall states(*8*), and other non-Fermi liquid behavior(*9*).

**Main Text:**

Anisotropy is a common property of materials and has profound effects on electronic structure and possible quantum phases. Some of the most striking examples are in layered van der Waals materials, where in-plane vs. out-of-plane anisotropy creates quasi-two-dimensional (quasi-2D) electronic states from three-dimensional (3D) crystals(*10*, *11*). Quasi-2D physics in general gives rise to many unusual quantum phenomena, such as high temperature superconductivity in cuprates(*12*, *13*), Dirac electrons(*14*), and new topological phases(*15*, *16*) in monolayer and rhombohedral graphene, and an indirect-to-direct semiconductor bandgap in monolayer transition metal dichalcogenides (TMDs)[16]. Anisotropy within two-dimensional materials can also create novel quantum electronic phases. An example is the interplay between strong Coulomb interactions and anisotropic electron mass that is predicted to give rise to oblique Wigner crystals(*3*, *17*) with antiferromagnetic order(*17*) at low electron density. At increased electron density, the quantum melting of such materials is expected to be strongly anisotropic and may lead to quasi-1D electron liquid crystal phases. Previous studies of hole-doped GaAs have revealed anisotropic electrical transport at low electron density(*18–21*), consistent with the presence of anisotropic Wigner crystals. Direct imaging of an oblique Wigner crystal and its quantum melting behavior, however, has never been performed.

Atomically thin 2D vdW heterostructures provide an ideal platform for directly imaging Wigner crystal physics facilitated by strong electron-electron interactions and accessibility of the 2D surface. This has enabled scanning tunneling microscopy-based imaging of generalized Wigner crystals in $WS_2$/$WSe_2$ moiré superlattices(*22*), Wigner molecular crystals in moiré heterostructures(*23*), 2D Wigner crystals under high magnetic field in bilayer graphene(*24*), and quantum melting of 2D Wigner solids at zero magnetic field in bilayer $MoSe_2$(*25*). Electrons in graphene and $MoSe_2$ have isotropic electron masses that lead to the formation of highly symmetric hexagonal Wigner crystals. Here we report the visualization of a two-dimensional oblique Wigner crystal and its quantum melting in triclinic monolayer rhenium diselenide ($ReSe_2$), a material exhibiting intrinsic electron effective-mass anisotropy, by using non-invasive scanning tunnelling microscopy (STM). At low electron density individual electrons in monolayer $ReSe_2$ are well-separated and exhibit an unusual elliptical density distribution. The orientation and aspect ratio of the elliptical charge density profile allow the first direct determination of the electron mass anisotropy direction and magnitude from STM imaging. We show that anisotropic electrons form an oblique Wigner crystal characterized by primitive lattice vectors of different lengths. At higher electron density, the oblique Wigner crystal exhibits anisotropic quantum melting with quantum fluctuations increasing much faster along the light-mass direction. Such quasi-1D melting of oblique Wigner crystals in $ReSe_2$ gives rise to an electronic smectic liquid crystal phase.

**STM topography of anisotropic $ReSe_2$ monolayer**

$ReSe_2$ has a stable 1T' phase characterized by a distorted 1T structure with a triclinic symmetry. Monolayer $ReSe_2$ has a pronounced in-plane effective mass anisotropy(*26–29*). Figure 1A illustrates our experimental setup and device configuration for characterizing anisotropic electron behavior in $ReSe_2$. The $ReSe_2$ monolayer is placed on top of an hBN dielectric layer and a graphite back gate. The electron density of the $ReSe_2$ layer is precisely tuned by the bottom gate voltage $V_{BG}$ between the graphite bottom gate and the $ReSe_2$. To achieve reduced contact resistance and efficient gating, graphene nanoribbon (GNR) arrays are employed as electrical contacts to the $ReSe_2$(*23*, *25*, *30*). Figure 1B displays an optical micrograph of the fabricated heterostructure device, showing the $ReSe_2$, GNR, and BG regions delineated by blue, yellow, and red solid lines, respectively.

Figure 1C illustrates the atomic structure of monolayer 1T'-ReSe2. The 2D lattice has no inversion or reflection symmetry. And the rhenium atoms from a chain structure along the b direction. Figure 1D displays an STM topographic image of the monolayer $ReSe_2$ for a 10nm×10nm region at a sample bias of $V_{bias} = -3V$, a tunnel current of $I_{sp} = 100pA$, and a backgate voltage of $V_{BG} = -8V$. The atomic lattice of the monolayer $MoSe_2$ is clearly resolved and exhibit a lattice constant of 6.6 Å. Three crystallographic directions of $ReSe_2$ are shown in the top-right corner, where the orange line indicates the rhenium chain direction, which traverses both the highest and lowest features in the topography(*28*). The inset to Figure 1D displays the Fast Fourier transform (FFT) of the STM topography in Fig. 1D. The FFT diffraction peaks corresponding to directions perpendicular to the orange rhenium chain have weaker intensity.

**Imaging oblique 2D Wigner crystals**

We employed a non-invasive conduction band edge (CBE) tunnel current measurement technique to directly probe Wigner crystals at low electron density with negligible perturbation(*25*, *30*). Figures 2A and 2B show the CBE tunnel current maps of Wigner crystal states obtained from two different regions of monolayer $ReSe_2$ for an electron density of $3.1 \times 10^{12}$cm$^2$ (see SI section 2 for electron density estimate; charged defects are indicated by red dashed circles in the images). Well-isolated individual electrons are observed in the charge density map. Each electron exhibits a highly anisotropic charge density profile with its elongated axis near to the rhenium chain direction (as denoted by the orange line in the top-right corner). The electrons form a Wigner crystal (WC) lattice with some distortion from charge defects. Figures 2C and 2D display the corresponding FFT power spectrum for the electron density maps in Figs. 2A and 2B, respectively and show clearly defined diffraction peaks.

We quantitatively analyze the anisotropic WC electron density by fitting each individual electron profile with a 2D Gaussian of the form:

$$I = A_0 + A_1 \exp\left\{-\left[\frac{(x-x_0)\cdot cos\theta+(y-y_0)\cdot sin\theta}{w_s}\right]^2 - \left[\frac{-(x-x_0)\cdot sin\theta+(y-y_0)\cdot cos\theta}{w_l}\right]^2\right\} \quad \text{(I)}$$

where $w_l$ and $w_s$ denote the widths along the long and short axes of the elliptical profile (as illustrated in Fig. 2H) and θ is the orientation of the long axis with respect to direction of minimal effective mass. Figures 2E and 2F show the profiles (dark blue dashed ellipses) where electron density drops to half its peak value for Figs. 2A, and 2B. The average electron aspect ratio ($w_l/w_s$) is 1.7$\pm$0.1 and the average orientation of the electron profile long axis is marked by green solid lines in Figs. 2E, and 2F. Interestingly, the long axes of the electron profiles are aligned close to (but not precisely) the rhenium chain direction of the atomic lattice (orange lines in upper right corner of Figs. 2E, and 2F).

An explanation for the elongated electron profile can be found by considering electrons with an anisotropic effective mass confined to an isotropic 2D harmonic potential V=1/2kr$^2$. The electron in the ground state will be more delocalized along the light mass direction, leading to an electron density of the form of eq.(I) with $w_l = \hbar^{1/2}k^{-1/4}{m_l}^{-1/4}$ and $w_s = \hbar^{1/2}k^{-1/4}{m_h}^{-1/4}$, where $m_h$ and $m_l$ are the heavy and light electron masses. Comparison of this model to STM data enables direct determination of the electron effective mass anisotropy in a 2D material from STM images. The extracted anisotropy ratio that we obtain for $ReSe_2$ is $m_h/m_l$ = $(w_l/w_s)^4$ = 8.4$\pm$2.0, where $w_l$ and $w_s$ are the fitted widths of the electron density profile and the light-mass direction is 9° $\pm$ 3.0° away from the rhenium chain direction. We can compare these experimental results with the electronic structure of monolayer $ReSe_2$ predicted by ab initio density functional theory (DFT). Figure 2G shows the energy-momentum contour near the conduction band minimum from our DFT calculation. It yields an electron mass anisotropy of $m_h/m_l$ = 6.45 and a light-mass axis (long white arrow) offset by 7.4° from the rhenium chain direction (dashed orange line), in reasonable agreement with our experimental observations.

We next investigated the lattice structure of the Wigner crystal. In Figs. 2E and 2F, the red, orange, and blue lines connect each electron to its nearest neighbors. Most electrons have six nearest neighbors, consistent with a locally ordered configuration. Despite the presence of some strain and disorder, the lattice can be characterized as an oblique 2D crystal with primitive vectors of unequal lengths. Figure 2H illustrates the real-space configuration of an ideal oblique Wigner crystal formed by anisotropic electrons. The shortest primitive vector ($\vec{a}_1$) lies approximately perpendicular to the elongated direction of the electron profile (green solid line in Figs. 2E and 2F), while the other two primitive vectors ($\vec{a}_2$, $\vec{a}_3$) are of equal length and have angles larger than 60° relative to $\vec{a}_1$. By averaging the unit cells in the experimental data, we obtain an $a_2/a_1$ length ratio of $1.13\pm0.01$. This anisotropy can also be observed in the FFT power spectra in Fig. 2C and 2D. In contrast to the hexagonal Wigner crystal expected for isotropic electrons, this oblique configuration reduces the Coulomb energy for anisotropic electrons. Our theory predicts an anisotropic Wigner crystal with an aspect ratio $\frac{a_2}{a_1} \approx 1.15$ across the experimental range $n > n_0 = 10^{12}\text{cm}^{-2}$, consistent with our experimental observations (see SI section 3 for details).

**Quantum melting of the oblique Wigner crystal**

We next investigated how oblique Wigner crystals evolve with increased electron density. Figures 3A, 3C, 3E, and 3G show the CBE tunnel current maps of the same region measured at $V_{BG}$=6V, 8V, 10V, and 12V, corresponding to electron densities of $3.1 \times 10^{12}$, $4.3 \times 10^{12}$, $5.4 \times 10^{12}$, and $6.6 \times 10^{12}\ cm^{-2}$, respectively (based on a parallel-plate capacitance model; see supplementary information for details). The light-mass direction is denoted by a green line in the upper left corner of Figure 3A and charged defects are labeled by red dashed circles in Figure 3A. Figures 3B, 3D, 3F, and 3H display the FFT power spectrum obtained from the tunnel current maps in Figs 3A, 3C, 3E, and 3G, respectively. At the lowest density of $3.1 \times 10^{12}\ cm^{-2}$ (Fig. 3A), the electrons form a local oblique Wigner crystal even in the presence of disorders. At the increased electron density of $4.3 \times 10^{12}\ cm^{-2}$ (Fig. 3C) quantum fluctuations are seen to increase faster along the light-mass direction, causing individual electrons to become less well defined along this direction in real space. This is accompanied by a small rotation of the diffraction spots (Fig. 3D). This trend continues at the higher electron density of $5.4 \times 10^{12}\ cm^{-2}$ (Fig. 3E) where individual electrons become even harder to discern along the light mass direction. At the electron density of $6.6 \times 10^{12}\ cm^{-2}$ (Fig. 3G) the electron density is almost homogeneous along the light-mass axis away from defects. This indicates that the WC has melted along this direction. In contrast, pronounced periodic modulation persists in the heavy-mass direction. The FFT power spectrum at the highest electron density (Fig. 3H) reveals only two bright peaks along the heavy-mass direction, consistent with the presence of 1D periodic lines in real space. Such anisotropic melting signifies the emergence of a quantum smectic phase - a partially melted Wigner crystal that retains translational order along one direction while becoming fluid-like along the orthogonal direction.

We can define an effective anisotropic Wigner-Seitz radius as: $r_{sl} = \frac{m_l e^2}{4\pi\varepsilon\hbar^2\sqrt{\pi n}}$ and $r_{sh} = \frac{m_h e^2}{4\pi\varepsilon\hbar^2\sqrt{\pi n}}$, $\varepsilon$ is the effective dielectric constant and *n* is the electron density. Using the theoretically predicted values of $m_l = 0.55 m_e$ and $m_h = 3.52 m_e$ and an effective dielectric constant of $\varepsilon \approx \frac{1}{2}\varepsilon_{hBN} + \frac{1}{2}\varepsilon_{vacuum} \approx 2.6$, the quasi-1d melted state at $n = 6.6 \times 10^{12} cm^{-2}$ correspond to $r_{sl}$ = 7.7 and $r_{sh}$ = 49.5. Such 1D melting behavior for our oblique WC is strikingly different from hexagonal WCs formed by isotropic electrons which tend to melt isotropically and form percolative liquid regions(*25*). For the anisotropic case melting is strongly directional and leads to a smectic-like intermediate phase.

Figures 4**A**-**C** show quasi-1D melting of the oblique WC in several different regions at an electron density of $7.1 \times 10^{12}\ cm^{-2}$(for Fig. 4**A** and **C**) and $6.6 \times 10^{12}\ cm^{-2}$(for Fig. 4**B**) with corresponding FFT power spectrum in Figs. 4**D**-**F**. The general melting behavior is universal in all regions, although local disorder can affect the detailed electron distributions. Each FFT power spectrum exhibits two pronounced diffraction peaks as expected for a smectic liquid crystal phase. The smectic melting direction is generally aligned with the light-mass direction. A small

deviation between these two directions can be caused by local disorder, which can distort or partially pin the electron configuration near defects.

In conclusion, we have visualized oblique Wigner crystals formed by anisotropic electrons and determined the electron mass anisotropy directly from STM images. We further observed the quasi-1D quantum melting of anisotropy WCs into electronic smectic liquid crystal phases. The smectic liquid crystal represents a novel coupled 1D electron system. It could give rise to exotic correlated physics, such as sliding Luttinger liquid(*8*, *31*, *32*) and new fractional quantum Hall states(*8*, *33*, *34*).

## Acknowledgements:

**Funding:** This work was primarily funded by the U.S. Department of Energy, Office of Science, Basic Energy Sciences, Materials Sciences and Engineering Division under Contract No. DE-AC02-05-CH11231 within the van der Waals heterostructure program KCFWF16 (device fabrication, STM spectroscopy, theoretical analysis and computations). Support was also provided by the Department of Defense through a Vannevar Bush Faculty Fellowship grant N00014-23-1-2869(surface preparation). This research used the Lawrencium computational cluster provided by the Lawrence Berkeley National Laboratory (Supported by the U.S. Department of Energy, Office of Basic Energy Sciences under Contract No. DE-AC02-05-CH11231) and resources of the National Energy Research Scientific Computing Center (NERSC), a U.S. Department of Energy Office of Science User Facility located at Lawrence Berkeley National Laboratory, operated under Contract No. DE-AC02-05CH11231 (DFT calculations). K.W. and T.T. acknowledge support from the JSPS KAKENHI (Grant Numbers 21H05233 and 23H02052), the CREST (JPMJCR24A5), JST and World Premier International Research Center Initiative (WPI), MEXT, Japan (Preparation of hBN crystals).

**Author contributions:**

Conceive the project: M.F.C., and F.W.

Sample fabrication: Z.X., and J.X.

STM/STS measurement: Z.X., and J.X.

DFT calculation: W.K., and S.G.L.

Experimental design and data analysis: Z.X., J.X., H.L., M.F.C. and F.W.

hBN crystal growth: T.T., and K.W.

Writing and review: all authors

**Competing interests:**

The authors declare no financial competing interests.

**Data and materials availability:**

The data supporting the findings of this study are included in the main text and in the Supplementary Information files, and are also available from the corresponding authors upon request.

**Supplementary Materials**

Materials and Methods

Supplementary Text

Figs. S1 to S2

References (1-40)

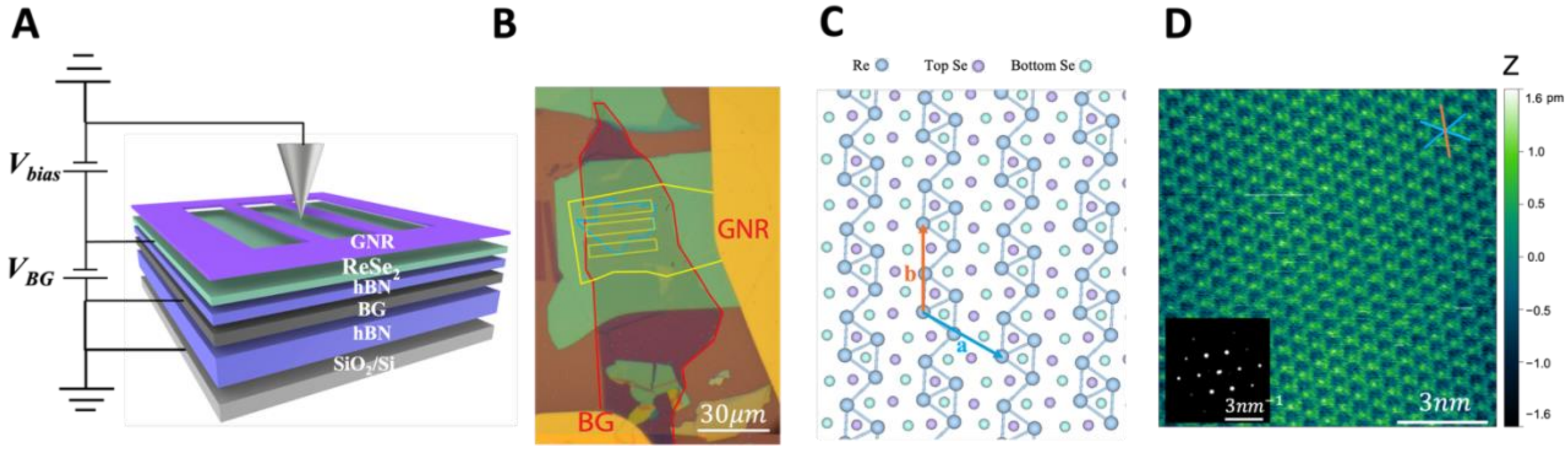

**Figure 1. STM measurement of monolayer $ReSe_2$. A**. Sketch of the STM measurement setup for a gate-tunable monolayer of $ReSe_2$. The $ReSe_2$ is placed on top of a 40nm thick hexagonal boron nitride (hBN) layer and a graphite back gate (BG). A back gate voltage $V_{BG}$ is applied to control the charge carrier density of the $ReSe_2$. A bias voltage $V_{bias}$ is applied between $ReSe_2$ and the STM tip to induce tunnel current. A graphene nanoribbon (GNR) array is placed on top of the $ReSe_2$ as a contact electrode. **B**. Optical microscopy image of the heterostructure. The $ReSe_2$, GNR, and BG regions are outlined in blue, yellow, and red, respectively. **C**. Illustration of $ReSe_2$ atomic structure. The Re, top Se, and bottom Se atoms are represented by light blue, light purple, and light cyan spheres, respectively. The rhenium chain structure breaks hexagonal symmetry in $ReSe_2$ and aligns with $\vec{b}$. **D**. STM topographic image of a $ReSe_2$ monolayer over a 10nm×10nm area, showing atomic resolution ($V_{bias} = -3V, I_{sp} = 100pA, V_{BG} = -8V$). The measured lattice constant along $\vec{a}$ and $\vec{b}$ directions are similar, both at around 6.6 Å. Top-right corner: Three solid lines indicate $ReSe_2$ lattice directions. The orange line marks the rhenium chain direction, which traverses both the highest and lowest features in the topography(*28*). Inset: Fast Fourier transform (FFT) of the STM image in (**D**).

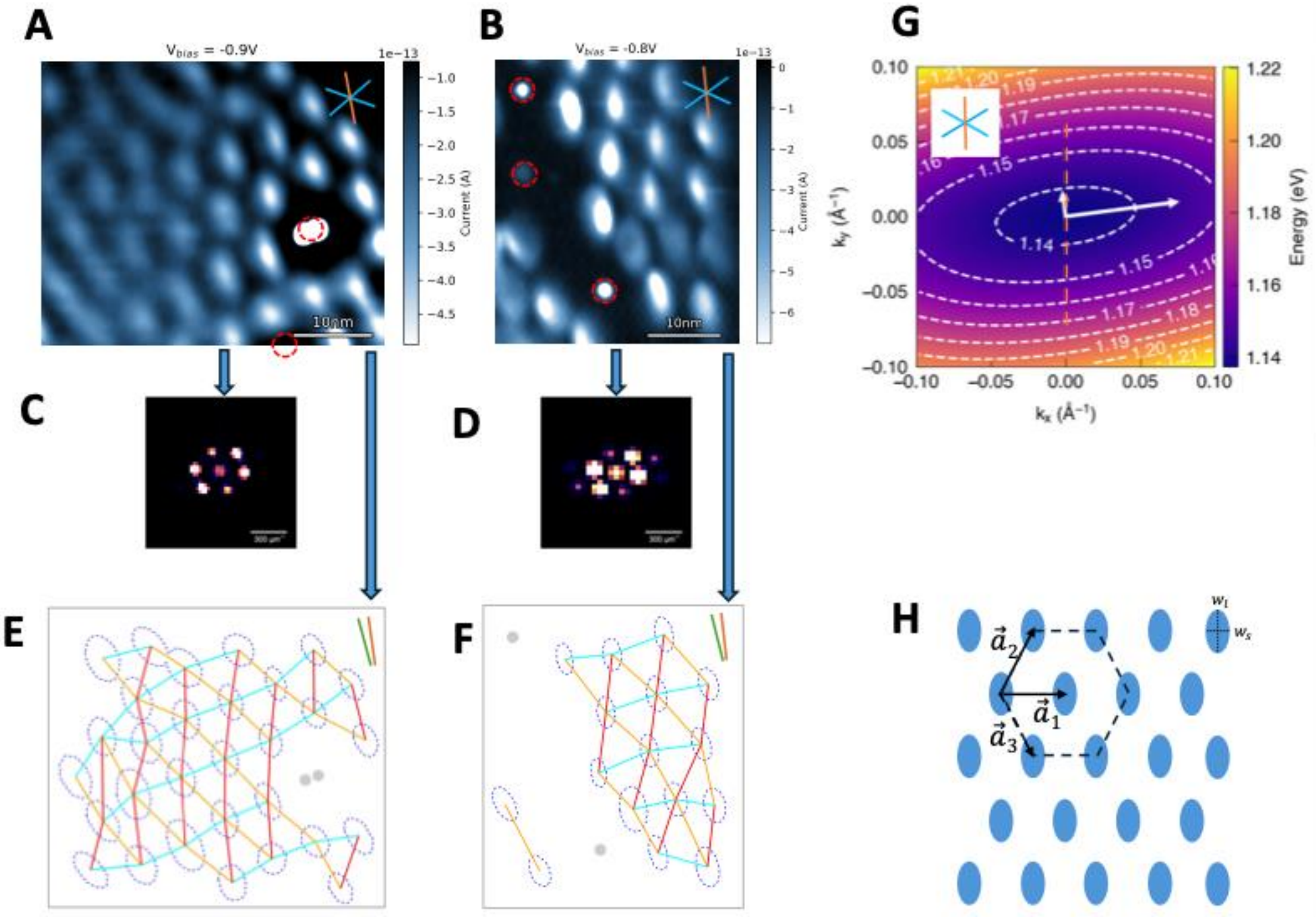

**Figure 2. Imaging anisotropic effects in 2D Wigner crystals. A, B**. CBE tunnel current maps of oblique Wigner crystals in two different regions having the same electron density of $3.1 \times 10^{12} cm^{-2}$. **A**, $V_{BG}$ = 6V, $V_{bias}$ = -0.9V. **B**, $V_{BG}$ = 6V, $V_{bias}$ = -0.8V. Bright spots correspond to individual electrons forming a Wigner crystal (WC). Charged defects are labeled by red dashed circles. The three solid lines in the top-right corner indicate lattice directions of $ReSe_2$, with the orange line denoting the rhenium chain direction. Each individual electron density profile exhibits an anisotropic elliptical shape. **C, D**. The FFT power spectra of (**A**) and (**B**), respectively, show clear diffraction spots from the anisotropic WC lattices. **E, F**. WC structure analysis of (**A**) and (**B**), respectively. The electrons are fitted using 2D Gaussian functions resulting in blue dashed ellipses showing the anisotropic FWHM having width of $w_l$ and $w_s$ along the major and minor axes, respectively. The centers of electrons are connected by solid lines in different colors, representing distinct orientations. The average measured $w_l/w_s$ ratio is $1.7 \pm 0.1$ and corresponds to an electron mass anisotropy ratio $m_h/m_l = (w_l/w_s)^4 \sim 8.4 \pm 2.0$. The green solid lines show the average light mass direction, which is tilted from the rhenium chain direction (orange line) by ~ $9° \pm 3°$. **G**. Conduction band structure of monolayer $ReSe_2$ near the conduction band minimum calculated by DFT predicts an electron mass anisotropy ratio $m_h/m_l = (w_l/w_s)^4 = 6.5$ and a 7.4° offset between the light-mass axis (long white arrow) and the rhenium chain direction (dashed orange line). **H**. Illustration of the anisotropic Wigner crystal configuration. Black arrows represent the primitive lattice vectors.

The electron shape anisotropy is characterized by the axes $w_l$ and $w_s$. The primitive lattice vector ratio $|\overrightarrow{a_2}|/|\overrightarrow{a_1}|$ is around 1.13 in the experimental data.

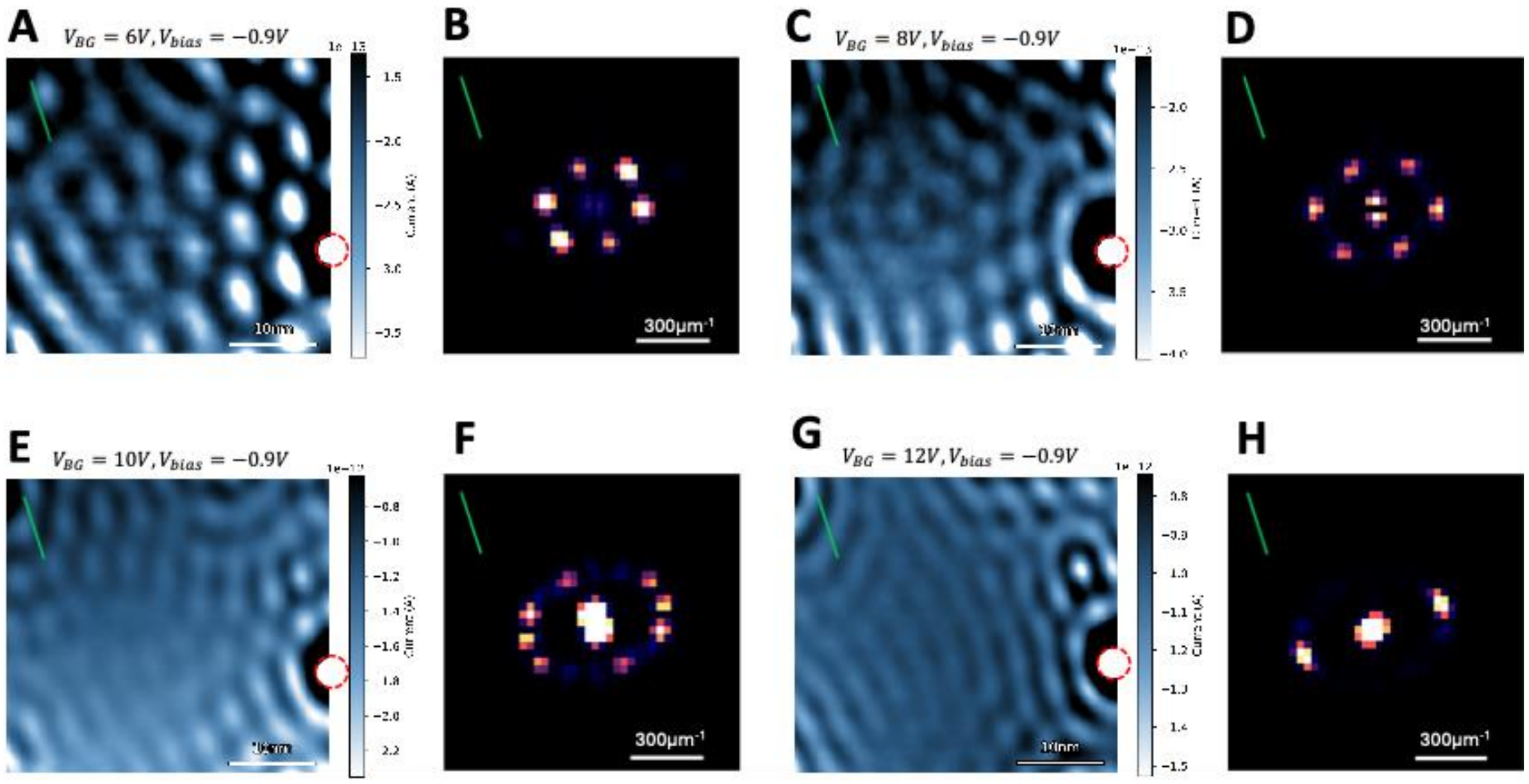

**Figure 3. Quantum melting of the anisotropic Wigner crystal. A, C, E, and G**. CBE tunnel current maps show evolution from (**A**) anisotropic Wigner crystal to (**G**) smectic liquid crystal as the electron density increases with increased $V_{BG}$. Charged defects are marked by red dashed circles in **A, C, E, and G**. $V_{bias} = -0.9V$: **A**, $V_{BG} = 6V$ $(n = 3.1 \times 10^{12} cm^{-2})$. **C**, $V_{BG} = 8V, (n = 4.3 \times 10^{12} cm^{-2})$. **E**, $V_{BG} = 10V, (n = 5.4 \times 10^{12} cm^{-2})$. **G**, $V_{BG} = 12V, (n = 6.6 \times 10^{12} cm^{-2})$. **B, D, F, and H**. The FFT power spectrum of **A, C, E, and G**, correspondingly.

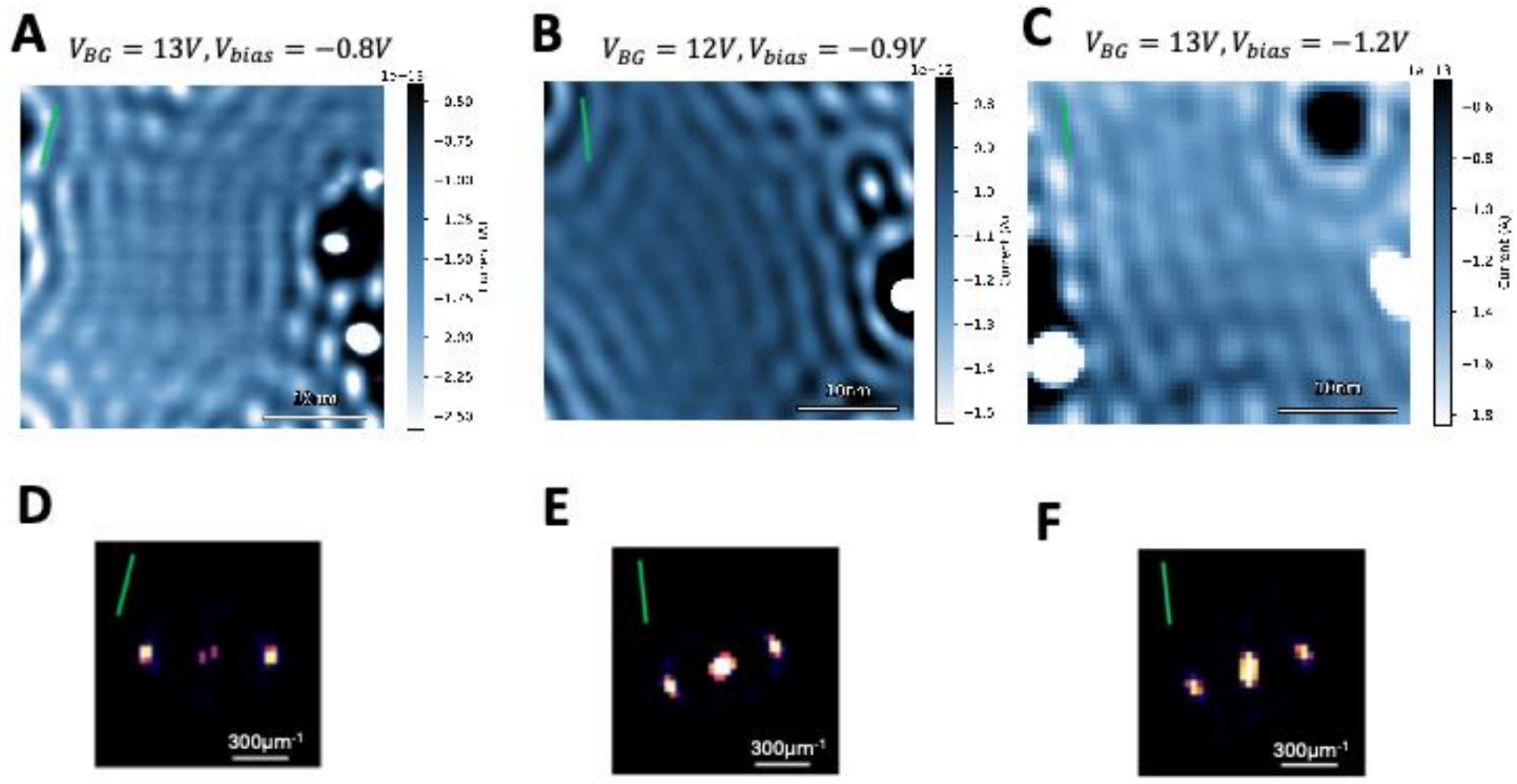

**Figure 4. Smectic liquid crystal phase of quasi-1D melted anisotropic electrons. A-C**. CBE tunnel current maps show the smectic liquid crystal phase of anisotropic electrons in three different regions at high electron density. **A**, $V_{BG}$ = 13V, $V_{bias}$ = -0.8V. **B**, $V_{BG}$ = 12V, $V_{bias}$ = -0.9V. **C**, $V_{BG}$ = 13V, $V_{bias}$ = -1.2V. **D-F**. The FFT power spectra corresponding to **A**-**C**, respectively. Power spectra signify quasi-1D melting and formation of smectic liquid crystal phase, light-mass direction shown by solid green line.